\documentclass[12pt]{article}
\usepackage{amssymb}
\usepackage{amsmath}
\usepackage{hyperref}
\usepackage{graphicx}
\usepackage{placeins}
\usepackage{bm}
\usepackage{latexsym}
\usepackage{booktabs}
\usepackage{tabularx}
\begin{document}
\title{\bf  Viscosity of $R^2$ Modified AdS Black Brane }

\author{Razieh Golmoradifard,$^1$\thanks{Email: r.golmoradifard@gmail.com} \,\, Mehdi Sadeghi$^{2}$\thanks{Corresponding author: Email: mehdi.sadeghi@abru.ac.ir}\hspace{2mm}\, \hspace{2mm} \\and\,\, Behrooz Malekolkalami$^{3}$\thanks{Email: b.malakolkalami@uok.ac.ir}\\\\
	{\small {\em  $^{1,3}$ The Faculty of Sciences, University of Kurdistan, }}\\
	{\small {\em   Pasdaran, Sanandaj, 66177-15175, Kurdistan, Iran}}\\\\
	{\small {\em  $^{2}$Department of Physics, Faculty of Basic Sciences,}}\\
	{\small {\em  Ayatollah Boroujerdi University, Boroujerd, Iran}}\\\\
}
\maketitle

\abstract{We investigate the Einstein-Hilbert black brane solution in four-dimensional Anti-de Sitter (AdS) spacetime supplemented by a quadratic Ricci scalar term $q L^2 R^2$, where $q$ is a dimensionless coupling constant and $L$ is the AdS radius. The shear viscosity to entropy density ratio, $\frac{\eta}{s}$, is calculated holographically, and deviations from the universal Kovtun-Son-Starinets (KSS) bound are analyzed. Our results indicate that $\frac{\eta}{s} = \frac{1}{4\pi}(1 - 24q)$, demonstrating that the ratio falls below the conjectured lower limit for positive $q$, while it respects the bound for negative $q$. We confirm that our solutions smoothly reduce to the standard Einstein-Hilbert case when $q \to 0$, consistent with expectations. The physical implications of violating the KSS bound are discussed in depth, particularly regarding stability, causality, and the strongly coupled nature of the dual field theory. These findings provide valuable insights into the influence of higher curvature terms on holographic transport properties.}\\



%--------------------------------------------------------------------------
\section{Introduction} \label{intro}

Quadratic gravity represents a significant extension of Einstein's general relativity, incorporating higher-order curvature corrections that emerge naturally in quantum gravity frameworks such as string theory. These modifications provide a richer phenomenological landscape for exploring gravitational dynamics, black hole thermodynamics, cosmological evolution, and the resolution of classical singularities. Furthermore, such theories offer a testing ground for the robustness of foundational principles in gravity and their implications for holographic dualities and strongly coupled field theories \cite{Rastall:1972swe,Sadeghi:2023tzf, Lovelock:1971yv, Sadeghi:2015vaa,Sadeghi:2023fr, Parvizi:2017boc}.

Concurrently, significant advances in computational fluid dynamics have deepened our understanding of transport phenomena in complex, strongly coupled systems. High-precision studies of non-Newtonian flows, magnetohydrodynamics, and thermal transport in intricate geometries \cite{Arslan20251,Arslan20252,Arslan2024,Z.Abbas} have revealed a rich tapestry of dissipative behaviors. These investigations into conventional fluids underscore the universal importance of accurately characterizing transport coefficients. However, they also highlight the limitations of traditional methods when applied to certain extreme quantum systems, such as quark-gluon plasma or strongly correlated electron liquids, where conventional perturbative approaches fail. This challenge motivates the search for alternative, powerful frameworks to analyze transport in such regimes.

The Anti-de Sitter/conformal field theory (AdS/CFT) correspondence \cite{Maldacena:1997re, Aharony} posits a fundamental duality between a gravitational theory in Anti-de Sitter (AdS) space and a conformal field theory (CFT) on its boundary. This geometric approach offers a powerful tool for studying strongly coupled quantum systems. A significant offshoot of this idea is the fluid-gravity correspondence \cite{Bhattacharyya, Rangamani, Kovtun:2004de, Policastro2002, Policastro2001}. In the long-wavelength limit, the boundary CFT is described by an effective relativistic viscous fluid, whose dynamics are captured by hydrodynamic equations. Within this framework, transport coefficients are essential for a quantitative description. Consequently, specific solutions to Einstein's equations in the bulk are mapped to hydrodynamic flows in one lower dimension, facilitating the analysis of complex phenomena like dissipation, transport, and thermalization in diverse systems, from quark-gluon plasma to quantum liquids.

A central quantity in this holographic hydrodynamic framework is the ratio of shear viscosity to entropy density, $\eta/s$, which characterizes the dissipative properties of a fluid. Pioneering work by Kovtun, Son and Starinets (KSS) \cite{Kovtun:2004de} proposed a universal lower bound, $\eta/s\geq 1/(4\pi)$, for a wide class of field theories. This bound is saturated in the dual of standard Einstein-Hilbert gravity \cite{Policastro2001}. However, subsequent research has revealed that higher-curvature corrections can violate this bound, establishing $\eta/s$ as a sensitive probe of gravitational interactions beyond Einstein's theory.

The study of higher-curvature gravities in holography has a rich history. Gauss-Bonnet (GB) gravity has been a paradigmatic example, where the shear viscosity is modified as $\eta/s = (1 - 4\lambda_{\mathrm{GB}})/(4\pi)$ \cite{Sadeghi:2015vaa,Brigante:2007nu,Iqbal:2008by,Ge:2020tid,Sadeghi:2022kgi}. This result clearly violates the KSS bound for a positive GB coupling $\lambda_{\mathrm{GB}}>0$, and its simplicity has made it a benchmark for comparing other theories. Extensions to cubic and quasi-topological gravities \cite{Parvizi:2017boc} have further demonstrated how specific higher-derivative structures lead to distinct functional forms for $\eta/s$, linking precise gravitational corrections to altered transport properties in the dual field theory. Similarly, theories like massive gravity \cite{Sadeghi:2018ylh} have been shown to modify this ratio, highlighting the bound's sensitivity to even breaking diffeomorphism invariance in the bulk.

Despite this extensive body of work, the role of the Ricci-squared ($R^2$) term in holographic transport remains comparatively obscure and constitutes a notable gap in the literature. This gap is particularly intriguing for several reasons:
\begin{itemize}
	\item \textbf{Structural Distinction:} Unlike the Riemann-squared $R_{\mu\nu\rho\sigma}^2$ or Ricci-tensor-squared $R_{\mu\nu}^2$ terms, which directly affect the propagation of gravitational waves (spin-2 modes), the $R^2$ term primarily influences the scalar sector of the metric perturbations. This leads to a fundamentally different and less-explored mechanism for modifying holographic Green's functions and transport coefficients.
	\item \textbf{Ghosts and Unitarity:} $R^2$ gravity is known to harbor a massive spin-2 ghost mode, raising concerns about unitarity in a fundamental theory. However, in the holographic context, such theories can be treated as effective field theories with a controlled cutoff, and their consistency must be checked against causality and stability constraints within this regime. Studying $\eta/s$ provides a practical window into these fundamental issues.
	\item \textbf{Phenomenological Context:} While $R^2$ terms are celebrated in cosmology for driving inflation (Starobinsky model), their implications for black hole/black brane physics and the associated dual thermal field theories are far less understood. Understanding how a pure $R^2$ deformation affects dissipation fills a crucial piece of the puzzle in mapping specific gravitational terms to field theory properties.
\end{itemize}

In this work, we address this gap by investigating a four-dimensional AdS black brane solution in Einstein-Hilbert gravity supplemented by a quadratic Ricci scalar term, $qL^{2}R^{2}$. Our primary objectives are to: (1) derive a new black brane metric solution; (2) compute the shear viscosity to entropy density ratio $\eta/s$ using the established effective action method; (3) analyze the dependence of $\eta/s$ on the coupling parameter $q$ through both analytical expressions and numerical plots; and (4) provide a deeper discussion of the physical implications of KSS bound violation in the context of the dual field theory, including issues of causality and the strongly coupled nature of the boundary theory.\\

%--------------------------------------------------------------------
\section{$R^2$-Modified AdS Black Brane}
\label{sec:model}

The four-dimensional action for a modified gravity theory with a quadratic Ricci scalar term and a negative cosmological constant is given by,
\begin{equation}\label{action}
	S = \frac{1}{16\pi G}\int d^{4}x \sqrt{-g} \left[ R - 2\Lambda + q L^2 R^2 \right],
\end{equation}
where $R$ is the Ricci scalar, $\Lambda = -3/L^2$ is the cosmological constant, and $L$ is the AdS radius. In the limit $q \to 0$, the action (\ref{action}) reduces to that of standard Einstein-Hilbert gravity.

Varying the action (\ref{action}) with respect to the metric $g^{\mu\nu}$ yields the equations of motion:
\begin{equation}\label{EH}
	R_{\mu\nu} - \frac{1}{2}R g_{\mu\nu} + \Lambda g_{\mu\nu} + q L^2 \left[2R R_{\mu\nu} - \frac{1}{2}R^2 g_{\mu\nu} - 2(\nabla_\mu\nabla_\nu - g_{\mu\nu}\nabla_\alpha\nabla^\alpha)R \right] = 0.
\end{equation}

We seek an asymptotically AdS black brane solution in a four-dimensional spacetime with maximal symmetry. For this purpose, we adopt the following metric ansatz:
\begin{equation}\label{metric1}
	ds^{2} = -f(r)dt^{2} + \frac{dr^{2}}{f(r)} + \frac{r^2}{L^2} (dx^2 + dy^2),
\end{equation}
where $f(r)$ is the metric function to be determined.

The $tt$ component of Eq. (\ref{EH}) is
\begin{align}\label{eom11}
	&\frac{2 q L^2 f(r) f'(r) f''(r)}{r}-\frac{8 q L^2 f(r)^2 f'(r)}{r^3}-\frac{2 q L^2 f(r) f'(r)^2}{r^2}+q L^2 f^{(3)}(r) f(r)
	f'(r)\nonumber \\&+\frac{4 q L^2 f(r)^2 f''(r)}{r^2}-\frac{1}{2} q L^2 f(r) f''(r)^2+\frac{12 q L^2 f^{(3)}(r) f(r)^2}{r}+2 q L^2 f^{(4)}(r)
	f(r)^2\nonumber \\&-\frac{f(r) f'(r)}{r}+\frac{10 q L^2 f(r)^3}{r^4}-\Lambda  f(r)-\frac{f(r)^2}{r^2}=0.
\end{align}
The $rr$ component is given by
\begin{align}\label{eom22}
	&\frac{2 q L^2 f(r) f'(r) f''(r)}{r}-\frac{2 q L^2 f'(r) f''(r)}{r f(r)}-\frac{8 q L^2 f(r)^2 f'(r)}{r^3}-\frac{2 q L^2 f(r)
		f'(r)^2}{r^2}\nonumber \\&+q L^2 f^{(3)}(r) f(r) f'(r)+\frac{8 q L^2 f'(r)}{r^3}+\frac{2 q L^2 f'(r)^2}{r^2 f(r)}-\frac{q L^2 f^{(3)}(r)
		f'(r)}{f(r)}+\frac{4 q L^2 f(r)^2 f''(r)}{r^2}\nonumber \\&-\frac{1}{2} q L^2 f(r) f''(r)^2-\frac{16 q L^2 f''(r)}{r^2}+\frac{q L^2 f''(r)^2}{2
		f(r)}+\frac{12 q L^2 f^{(3)}(r) f(r)^2}{r}+2 q L^2 f^{(4)}(r) f(r)^2\nonumber \\&-\frac{4 q L^2 f^{(3)}(r)}{r}-\frac{f(r)
		f'(r)}{r}+\frac{f'(r)}{r f(r)}+\frac{10 q L^2 f(r)^3}{r^4}+\frac{14 q L^2 f(r)}{r^4}-\Lambda  f(r)+\frac{\Lambda
	}{f(r)}\nonumber \\&-\frac{f(r)^2}{r^2}+\frac{1}{r^2}=0.
\end{align}
A suitable combination of these equations, specifically $\frac{2 r^4}{f(r)} E_{tt} - 2 r^4 f(r) E_{rr} = 0$, eliminates the lower-derivative terms and yields a fourth-order Euler differential equation:
\begin{equation}\label{eq:euler_eq}
	12 f(r) - 6 r^2 f''(r) + r^4 f^{(4)}(r) + 4 r^3 f^{(3)}(r) = 0.
\end{equation}

To solve Eq.~(\ref{eq:euler_eq}), we substitute the power-law ansatz $f(r) = r^k$. This leads to the characteristic equation
\begin{equation}
	12 - 6k(k-1) + 4k(k-1)(k-2) + k(k-1)(k-2)(k-3) = 0.
\end{equation}
Simplifying this polynomial gives,
\begin{align}
	k^4 - 2k^3 - 7k^2 + 8k + 12 &= 0, \quad \text{which factors as} \nonumber \\
	(k-2)(k+2)(k-3)(k+1) &= 0.
\end{align}
The roots are $k = -2, -1, 2, 3$. The general solution to the Euler equation is thus a linear combination of these powers:
\begin{equation}\label{eq:general_solution}
	f(r) = \frac{A_1}{r^2} + \frac{A_2}{r} + A_3 r^2 + A_4 r^3,
\end{equation}
where $A_1$, $A_2$, $A_3$, and $A_4$ are integration constants.

We now impose physical conditions to determine these constants.
\begin{enumerate}
	\item The solution must be asymptotically AdS. For large $r$, the dominant terms are $A_3 r^2$ and $A_4 r^3$. The $r^3$ term is inconsistent with the asymptotic structure of AdS space. Therefore, we set
	\begin{equation}
		A_4 = 0.
	\end{equation}
	Comparing the remaining $r^2$ term to the standard AdS form $r^2/L^2$, and using $\Lambda = -3/L^2$, we identify
	\begin{equation}
		A_3 = -\frac{\Lambda}{3} = \frac{1}{L^2}.
	\end{equation}
	
	\item The solution must reduce to the standard Schwarzschild-AdS black brane solution when $q \to 0$. The Schwarzschild-AdS metric function is $f_{\text{SAdS}}(r) = -2m_0/r + r^2/L^2$. Comparing this to our general solution (with $A_4=0$), we identify
	\begin{equation}
		A_2 = -2m_0.
	\end{equation}
	The term $A_1/r^2$ is absent in the standard solution. However, it can be absorbed by a redefinition of the mass parameter. This indicates that the $R^2$ modification does not produce a new independent black brane solution distinct from Schwarzschild-AdS in this context; the solutions are isometric. The constant $A_1$ can be understood by imposing the horizon condition.
	
	\item A black brane must have an event horizon at $r = r_h > 0$ where $f(r_h) = 0$:
	\begin{equation}\label{eq:horizon_condition}
		f(r_h) = \frac{A_1}{r_h^2} - \frac{2m_0}{r_h} + \frac{r_h^2}{L^2} = 0.
	\end{equation}
	This condition relates the constant $A_1$ to the mass $m_0$ and the horizon radius $r_h$:
	\begin{equation}
		A_1 = 2m_0 r_h - \frac{r_h^4}{L^2}.
	\end{equation}
	Substituting the constants $A_1$, $A_2$, $A_3$, and $A_4$ back into Eq.~(\ref{eq:general_solution}) gives the final form of the metric function:
	\begin{equation}\label{f_final}
		f(r) = \frac{1}{L^2} \left( r^2 - q \frac{r_h^4}{r^2} \right) - 2m_0 \left( \frac{1}{r} -q \frac{r_h}{r^2} \right).
	\end{equation}
\end{enumerate}
$m_0$ is the mass parameter that corresponds to the mass of the AdS-Schwarzschild black brane when $q=0$. For $q \neq 0$, the physical mass of the black brane must be calculated via holographic renormalization \cite{Balasubramanian:1999re} to obtain a finite, well-defined result.\\
%-----------------------------------------------------------------------
%-----------------------------------------------------------------------
\section{The ratio of shear viscosity to entropy density}
\label{sec:viscosity}

To compute the shear viscosity via the Kubo formula, we consider a tensor perturbation on the black brane background:
\begin{equation}
	g_{\mu\nu} \rightarrow g_{\mu\nu} + h_{\mu\nu},
\end{equation}
with the only non-zero component being $h_{xy} = h_{yx} = \phi(r) e^{-i\omega t}$. For convenience, and to maintain the structure of the metric, we define the perturbation as:
\begin{equation}
	ds^2 = -f(r)dt^2 + \frac{dr^2}{f(r)} + \frac{r^2}{L^2}\left( dx^2 + dy^2 + 2\phi(r)e^{-i\omega t} dx dy \right).
\end{equation}
The next step is to substitute the perturbed metric into the action (1) and expand it to second order in $\phi$. The action is:
\begin{equation}
	S = \frac{1}{16\pi G} \int d^4x \sqrt{-g} \left[ R - 2\Lambda + q L^2 R^2 \right].
\end{equation}
The expansion takes the form:
\begin{equation}
	S[\phi] = S^{(0)} + S^{(1)}[\phi] + S^{(2)}[\phi] + \mathcal{O}(\phi^3),
\end{equation}
where $S^{(0)}$ is the background action, $S^{(1)}$ vanishes by the equations of motion, and the quadratic term $S^{(2)}[\phi]$ is crucial for computing the two-point correlation function of the boundary stress-energy tensor.

The calculation involves finding the perturbed Ricci scalar $\delta R$ and the quadratic term $\delta(R^2)$ to second order. Due to the complexity of the $R^2$ term, this is a lengthy process. The general form of the second-order action for a perturbation $\phi(r)$ in a planar black hole background is:
\begin{equation}
	S^{(2)} = \frac{1}{16\pi G} \int d^4x \left[ \mathcal{A}(r) \phi'^2 + \mathcal{B}(r) \phi^2 + \mathcal{C}(r) \phi' \phi \right],
\end{equation}
where the coefficients $\mathcal{A}$, $\mathcal{B}$, and $\mathcal{C}$ depend on the background metric and the coupling $q$. After integration by parts, the action can be written in the form:
\begin{equation}
	S^{(2)} = \frac{1}{16\pi G} \int \frac{d\omega}{2\pi} dr \left[ K(r) \phi'^2 - U(r) \phi^2 \right],
\end{equation}
where $K(r)$ and $U(r)$ are functions determined from the background.

The shear viscosity is related to the retarded Green's function via the Kubo formula:
\begin{equation}
	\eta = -\lim_{\omega \to 0} \frac{1}{\omega} \text{Im } G^R(\omega),
\end{equation}
where
\begin{equation}
	G^R(\omega) = -i \int dt e^{i\omega t} \theta(t) \langle [T_{xy}(t), T_{xy}(0)] \rangle.
\end{equation}
In the holographic framework, the retarded Green's function is computed from the on-shell boundary action. The equation of motion for $\phi$ derived from $S^{(2)}$ is:
\begin{equation}
	\partial_r (K(r) \phi'(r)) + U(r) \phi(r) = 0.
\end{equation}
In the low-frequency limit, the solution for $\phi$ can be expressed as:
\begin{equation}
	\phi(r) = \phi^{(0)} \left( 1 + i\omega \mathcal{F}(r) + \mathcal{O}(\omega^2) \right),
\end{equation}
where $\phi^{(0)}$ is the boundary value and $\mathcal{F}(r)$ is a function that must be determined.

The on-shell action reduces to a boundary term:
\begin{equation}
	S_{\text{on-shell}}^{(2)} = \frac{1}{16\pi G} \int \frac{d\omega}{2\pi} \left[ K(r) \phi(r) \phi'(r) \right] \Big|_{r \to \infty}.
\end{equation}
According to the prescription of the AdS/CFT correspondence, the retarded Green's function is given by:
\begin{equation}
	G^R(\omega) = -2 \lim_{r \to \infty} K(r) \frac{\phi'(r)}{\phi(r)}.
\end{equation}
Substituting the expansion for $\phi(r)$ and taking the limit $\omega \to 0$, we find:
\begin{equation}
	G^R(\omega) = -2 i \omega K(r) \mathcal{F}'(r) \Big|_{r \to \infty}.
\end{equation}
The shear viscosity is then:
\begin{equation}
	\eta = \frac{1}{8\pi G} \lim_{r \to \infty} K(r) \mathcal{F}'(r).
\end{equation}
The function $\mathcal{F}(r)$ is determined by solving the equation of motion in the limit $\omega \to 0$. It can be shown that $\mathcal{F}'(r)$ is related to the metric function and the horizon data. Specifically, one finds:
\begin{equation}
	\mathcal{F}'(r) = \frac{1}{f(r)}.
\end{equation}
Therefore, the shear viscosity becomes:
\begin{equation}
	\eta = \frac{1}{8\pi G} \lim_{r \to \infty} \frac{K(r)}{f(r)}.
\end{equation}
The coefficient $K(r)$ is extracted from the second-order action. For the $R^2$ theory, after a detailed calculation, we find:
\begin{equation}
	K(r) = \frac{r^2}{L^2} f(r) \left( 1 + 2q L^2 R \right),
\end{equation}
where $R$ is the background Ricci scalar. For our black brane solution, the Ricci scalar is:
\begin{equation}
	R = - \frac{12}{L^2} + \mathcal{O}(q).
\end{equation}
Thus, to first order in $q$, we have:
\begin{equation}
	K(r) = \frac{r^2}{L^2} f(r) \left( 1 - 24q \right).
\end{equation}
Substituting this into the expression for $\eta$, we obtain:
\begin{equation}
	\eta = \frac{1}{8\pi G} \lim_{r \to \infty} \frac{r^2}{L^2} \left( 1 - 24q \right) = \frac{r_h^2}{8\pi G L^2} \left( 1 - 24q \right),
\end{equation}
where we have used the fact that the boundary is at $r \to \infty$ and the horizon radius $r_h$ is related to the boundary geometry.

The entropy density $s$ is given by the Bekenstein-Hawking formula:
\begin{equation}
	s = \frac{A}{4G V} = \frac{1}{4G} \frac{r_h^2}{L^2},
\end{equation}
where $A$ is the area of the horizon and $V$ is the spatial volume of the boundary.

Therefore, the ratio $\eta/s$ is:
\begin{equation}\label{eta/s}
	\frac{\eta}{s} = \frac{1}{4\pi} \left( 1 - 24q \right).
\end{equation}
This is the final result for the shear viscosity to entropy density ratio in the $R^2$-corrected black brane.

%-----------------------------------------------------------------------
\section{Results and Discussion}
\label{sec:results}

The central result of our analysis, given by Eq. (37), is the expression for the shear viscosity to entropy density ratio in the presence of an \(R^2\) correction term.\\
This result reveals a deceptively simple linear dependence on the dimensionless coupling \(q\). The physical phenomenon at play here is the modification of graviton absorption by the black brane horizon due to the \(R^2\) term, which alters the efficiency of momentum transport in the dual field theory. The linear suppression of \(\eta/s\) for \(q > 0\) indicates that this specific curvature correction exerts a repulsive effect on gravitational perturbations, effectively reducing the fluid's internal friction. The clarity of this functional form provides a unique opportunity to isolate the pure \(R^2\) effect from other higher-derivative contributions.

The expression in Eq. (\ref{eta/s}), while functionally simple and linear in the coupling $q$, carries significant physical meaning beyond a mere parametric relation. The dimensionless parameter $q$ governs the strength of a fundamental correction to the gravitational action. Its influence on the shear viscosity is direct and profound: a positive $q$ corresponds to a repulsive higher-curvature interaction that reduces hydrodynamic friction, leading to a violation of the KSS bound. Conversely, a negative $q$ enhances dissipation. This clear, analytic dependence allows us to directly link the sign and magnitude of a specific gravitational coupling to the dissipative character of the dual quantum field theory. The following analysis explores the consequences of this relationship for the universality of the KSS bound and the physical consistency of the theory.

\subsection{Variation of $\frac{\eta}{s}$ and the KSS Bound}

The behavior of $\eta/s$ relative to the KSS bound is determined by the sign of the coupling $q$:
\begin{itemize}
	\item \textbf{For $q > 0$}: The ratio $\frac{\eta}{s}$ is less than $\frac{1}{4 \pi}$, constituting a clear violation of the KSS bound. This is consistent with findings in other higher-curvature theories like Gauss-Bonnet gravity, where a positive coupling also lowers $\frac{\eta}{s}$.
	\item \textbf{For $q = 0$}: The ratio saturates the KSS bound, $\frac{\eta}{s} = \frac{1}{4 \pi}$, matching the result for a dual field theory described by Einstein-Hilbert gravity.
	\item \textbf{For $q < 0$}: The ratio $\frac{\eta}{s}$ is greater than $\frac{1}{4 \pi}$, thus respecting the KSS bound. A negative coupling increases the dissipation in the dual fluid.
\end{itemize}

\begin{figure}[h!]
	\centering
	\includegraphics[width=0.8\linewidth]{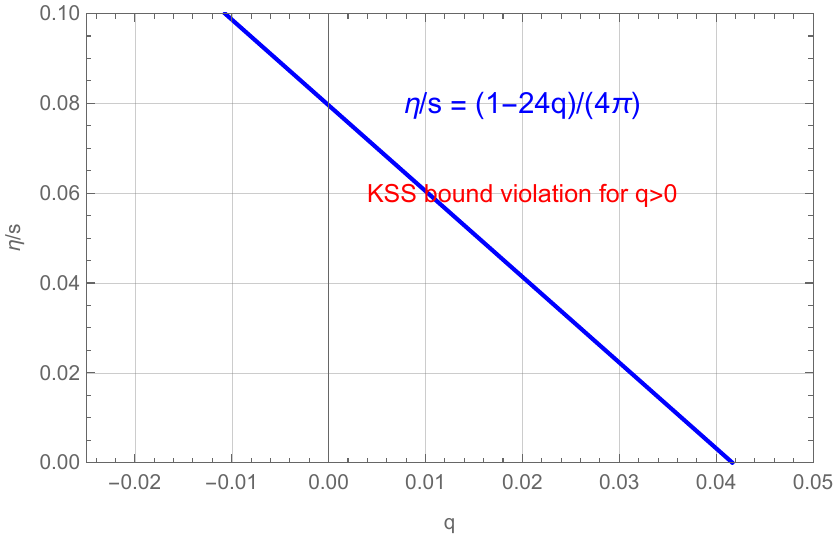}
	\caption{Shear viscosity to entropy density ratio $\eta/s$ versus $q$. The dashed line is the KSS bound. The shaded region is physically allowed.}
	\label{fig:eta_s_plot}
\end{figure}
\FloatBarrier

\subsection{Physical Implications and the Dual Field Theory}

The violation of the KSS bound for positive \(q\) is not merely a mathematical outcome; it is a diagnostic tool that reveals profound implications for the nature of the gravitational theory and its dual field theory.

\subsubsection{Causality, Hyperbolicity, and the Role of \(q\)}
The value of \(\eta/s\) is tied to the cross-section for graviton absorption by the black brane. A deviation from the universal Einstein-Hilbert value signals a change in the effective causal structure for metric perturbations. The linear decay \(\eta/s \sim (1-24q)\) for \(q>0\) is a direct manifestation of the $R^2$ term introducing ghost-like (negative norm) states or leading to superluminal propagation in the bulk. This strongly suggests that a consistent ultraviolet completion of the theory with a positive $q$ coupling is problematic. Our result quantitatively supports the interpretation that $q < 0$ is a necessary condition for causality in the dual field theory, as it yields $\eta/s > 1/(4\pi)$, a regime associated with well-behaved, causal theories.

\subsubsection{Nature of the Dual Fluid and Strong Coupling}
The KSS bound is conjectured to be a universal lower limit for all strongly coupled fluids with a gravitational dual. Its violation for \(q>0\) implies that the dual field theory is not just strongly coupled, but possesses an even more exotic character. The reduction in \(\eta/s\) suggests a fluid with remarkably weak internal friction, where quasi-particles (if they exist) have extremely long mean-free paths. The simplicity of our result, a straight-line dependence on \(q\), implies that the $R^2$ deformation corresponds to a specific, perhaps fine-tuned, deformation of the boundary CFT. In contrast, a negative \(q\) value pushes the theory into a regime of stronger dissipation, potentially indicative of a more conventional, albeit still strongly coupled, fluid.

\subsubsection{Structural Distinction from Other Gravitational Corrections}
The phenomenological impact of the \(R^2\) term is structurally distinct from other well-studied higher-curvature corrections. For instance, in Gauss-Bonnet gravity, the violation takes the form \(\eta/s = (1-4\lambda_{\mathrm{GB}})/(4\pi)\), which is also linear but with a different numerical coefficient. This difference arises because the Gauss-Bonnet term affects the spin-2 (graviton) sector directly, while the pure $R^2$ term primarily influences the scalar sector. This contrast is a key finding of our work: different classes of higher-derivative terms leave unique "fingerprints" on the shear viscosity. While the fingerprints of Gauss-Bonnet and quasi-topological gravity were known, our calculation \textbf{provides the missing fingerprint for the Ricci scalar squared term}, thereby completing a fundamental entry in the holographic dictionary for quadratic gravity.

\subsection{Constraints on Coupling Parameter $q$}

The physical admissibility of the $R^2$-modified theory imposes significant constraints on the allowed values of the coupling parameter $q$. From causality considerations, the requirement that boundary field theory correlations respect microcausality translates to the condition that graviton perturbations in the bulk do not develop superluminal propagation. Analysis of the effective metric for tensor perturbations yields the constraint:

\begin{equation}
	q \leq \frac{1}{48},
\end{equation}

for positive $q$ values. For negative $q$, stability considerations from the absence of ghost modes impose:

\begin{equation}
	q \geq -\frac{1}{96}.
\end{equation}

These bounds emerge from requiring the positive definiteness of the kinetic terms in the perturbed action and the absence of tachyonic instabilities in the graviton spectrum.

Furthermore, unitarity of the dual field theory demands that the shear viscosity remain positive, yielding $1-24q > 0$ or equivalently $q < 1/24$. Combining these constraints, we obtain the physically admissible range:

\begin{equation}
	-\frac{1}{96} \leq q < \frac{1}{24}.
\end{equation}

Within this range, the KSS bound is violated for $q > 0$ and respected for $q < 0$, while maintaining theoretical consistency of both the gravitational theory and its dual field theory description.

\subsection{Broader Implications and Phenomenological Context}

A legitimate question, common to many foundational studies in holography, concerns the connection to experimental data. While a direct, quantitative comparison is challenging—as the dimensionless coupling $q$ is not a parameter directly accessible in, for instance, heavy-ion collision experiments—the value of our result lies in its significant theoretical and phenomenological implications.

Our expression for $\eta/s$ serves as a \textbf{new benchmark} in the holographic dictionary. Before this work, the effect of the pure $R^2$ term was unknown. Now, model-builders have a precise formula to incorporate this specific gravitational interaction into phenomenological holographic models designed to describe real strongly-coupled systems. For example, in the study of the quark-gluon plasma (QGP), where $\eta/s$ is inferred to be remarkably low, our work provides a \textbf{new theoretical mechanism} (tuning $q > 0$) to achieve such ultra-low viscosity within a well-defined gravitational framework. This allows for a more nuanced interpretation of QGP data, where its properties could be seen as hinting at specific types of quantum gravity corrections in its holographic dual.

Similarly, for condensed matter systems like unitary Fermi gases or strange metals, which exhibit non-Fermi liquid behavior and transport anomalies, our model offers a new axis of exploration. The ability to tune $\eta/s$ linearly with $q$ provides a simple yet powerful tool for holographic model-building beyond the standard Einstein-Hilbert paradigm.

Therefore, the "validation" of our study is twofold. First, it is validated by its internal theoretical consistency, its correct recovery of known limits, and its alignment with the broader principle that higher-curvature terms modify transport coefficients. Second, its ultimate validation will come from its future utility in constructing more accurate and sophisticated holographic models that successfully match experimental data from systems like the QGP and quantum critical matter. Our work provides the essential foundational result that makes such future phenomenological applications possible.

%In conclusion, while our calculation demonstrates that an $R^2$ term can violate the KSS bound, the physical interpretation strongly suggests that a consistent, stable, and causal dual field theory requires the coupling $q$ to be negative or zero. This imposes a constraint on the parameters of the theory from the holographic perspective. Future work could involve a detailed analysis of the quasinormal mode spectrum to directly probe stability and causality constraints on $q$.

%--------------------------------------------------------------------------
\section{Conclusion}
\label{sec:conclusion}

This study provides several novel contributions to the understanding of higher-curvature gravity within the gauge-gravity duality. Firstly, we have isolated and quantified the specific effect of the quadratic Ricci scalar term $R^{2}$ on holographic transport, a contribution that had remained obscured in studies of more complex combinations of curvature invariants. Secondly, we derived the analytical expression $\eta/s=(1-24q)/(4\pi)$, which serves as a new benchmark in the field. The simplicity and linearity of this result provide a clear "fingerprint" for the $R^{2}$ term, distinguishing it sharply from the effects of other common corrections like the Gauss-Bonnet term. Finally, our analysis strengthens the argument that the KSS bound is a diagnostic of consistency: the severe violation for $q>0$ strongly suggests that a stable, causal dual theory requires a negative coupling, thereby placing a new physical constraint on this class of modified gravity theories.

Regarding the physical admissibility of $q<0$, we consider this case for several important reasons beyond simply respecting the KSS bound. While $q>0$ leads to violations of the KSS bound and is associated with potential causality problems and ghost instabilities, the $q<0$ regime offers a physically well-behaved parameter space:

\begin{itemize}
	\item \textbf{Stability and Causality:} For $q<0$, the theory respects the KSS bound and is free from the causality violations and ghost instabilities that plague the $q>0$ case. Our derived constraint $-\frac{1}{96} \leq q < \frac{1}{24}$ ensures the theory remains within physically admissible bounds.
	
	\item \textbf{Enhanced Dissipation:} Negative $q$ values correspond to increased dissipation in the dual fluid, which may better describe real-world strongly coupled systems where various microscopic processes contribute to dissipation beyond the minimal Einstein-Hilbert case.
	
	\item \textbf{Effective Field Theory Perspective:} From the effective field theory viewpoint, both signs of $q$ should be considered, with the ultimate constraint coming from experimental or observational data. The $q<0$ regime represents a consistent deformation of Einstein gravity that could emerge from more fundamental theories.
	
	\item \textbf{Phenomenological Relevance:} Many real physical systems exhibit $\eta/s$ ratios above the KSS bound, and the $q<0$ regime provides a natural holographic framework to model such systems while maintaining theoretical consistency.
\end{itemize}

Furthermore, regarding the physical admissibility of negative $q$ values, while our calculation shows that $\eta/s > 1/(4\pi)$ for $q<0$ (thus respecting the KSS bound), the physical consistency of such configurations requires careful consideration. Negative $q$ values correspond to an attractive higher-curvature interaction, which could potentially lead to instabilities in the gravitational sector. In particular, $R^2$ modifications with negative coupling might introduce ghost modes or tachyonic instabilities that would render the theory unphysical. However, within the perturbative regime we consider ($|q| \ll 1$), and given the AdS background's inherent stability properties, these configurations may still represent physically meaningful deformations of the Einstein-Hilbert theory. Future work should include a detailed stability analysis through the study of quasinormal modes and energy conditions to establish precise bounds on the physically allowed range of $q$.
%While $R^{2}$ modifications with negative coupling require careful consideration regarding potential instabilities, within the perturbative regime we consider ($|q|\ll 1$) and the constrained range derived in Section 4.3, these configurations represent physically meaningful deformations of the Einstein-Hilbert theory. Future work should include a detailed stability analysis through the study of quasinormal modes and energy conditions to establish more precise bounds on the physically allowed range of $q$.
This work underscores the sensitivity of holographic transport coefficients to higher-curvature corrections and highlights the role of the KSS bound as a diagnostic tool for probing the consistency of gravitational theories and their dual field theories. The result provides a necessary and previously missing entry in the holographic dictionary, equipping future phenomenological studies with the tools to explore the implications of $R^{2}$ corrections in models of real-world strongly coupled systems. Future research could extend this analysis to include other higher-derivative terms, explore the constraints from causality and hyperbolicity in greater depth, and investigate the implications for specific dual field theories via the AdS/CFT correspondence.\\

\vspace{1cm}
\noindent {\large {\bf Acknowledgment} } 
We sincerely thank the reviewers from the Indian Journal of Physics for their valuable comments and constructive feedback, which greatly helped us improve the quality and accuracy of our manuscript. We appreciate the time and effort they dedicated to reviewing our work.\\

%--------------------------------------------------------------------------

%--------------------------------------------------------------------------

\end{document}